\documentstyle{elsart}
\input epsf
\begin{document}
\begin{frontmatter}
\title{Front localization in a ballistic annihilation model.} 
\author{Pierre-Antoine Rey and Michel Droz\thanksref{MD}}
\address{D\'epartement de Physique Th\'eorique, Universit\'e de Gen\`eve,
CH-1211 Geneva 4, Switzerland}
\author{Jaros\l aw Piasecki\thanksref{THANK_JAREK}}
\address{Institute of Theoretical Physics, Warsaw University, Ho\.za 69, 
PL-00 681 Warsaw, Poland}

\thanks[MD]{Works partially supported by the Swiss National Science 
Foundation}
\thanks[THANK_JAREK]{Work partially supported by the Committee for Scientific 
Research (Poland), grant 2 P302 074 07.}
\vspace{10mm} 
{\centerline{\bf UGVA-DPT 1996/07-934}}
\vspace{10mm}

\begin{abstract}
We study the possibility of localization of the front present in a 
one-dimensional ballistically-controlled annihilation model in which the two 
annihilating species are initially spatially separated. We construct two 
different classes of initial conditions, for which the front remains localized.
\end{abstract}


\end{frontmatter}

\section{Introduction}

During the last decade, a large body of work has been devoted to the study of 
the kinetics of diffusion-annihilation processes. It is now well established 
that, below some upper critical dimension, the fluctuations play a central role 
and that accordingly mean-field like approximations are inappropriate.

Moreover, for the two species case $A+B \to 0$, when the two reactants are 
initially spatially separated, a reaction-diffusion front, getting larger with 
time, is formed~\cite{racz,cordro}. In the long time regime, the time dependent 
properties of this front (position, width) are characterized by power laws which 
are non-mean-field at or below two dimensions.

Instead of investigating a time dependent problem, it was shown by Cornell and 
Droz~\cite{prl_cor_droz} that it may be advantageous to study the front formed 
in the steady state reached by imposing antiparallel currents $J_A=|J|$ and 
$J_B=-|J|$ of A- and B-particles at $x=-\infty$ and $x=+\infty$ respectively. It 
turns out that first, exact prediction can be made in the one dimensional case 
and second this stationary problem is closely related to the time-dependent one.

More recently, a different but related problem has been considered, namely the
case of ballisti\-cally-controlled annihilation processes. Most of the results 
obtained are for the one dimensional case. Initially, the particles are randomly 
distributed in space and their velocities are distributed according to a given 
distribution. The particles move freely and when two of them collide they
instantaneously annihilate each other and disappear out of 
the system~\cite{ef,reddue,pia_uno,pia_due,pia_tre,poisson}. In 1985, Elskens 
and Frisch~\cite{ef} considered the case of one species of particles moving with 
velocities $+c$ and $-c$. The case of an arbitrary discrete velocity 
distribution has been studied by Droz, Rey, Frachebourg and 
Piasecki~\cite{pia_due,pia_tre} and exactly solved for a symmetric three 
velocities distribution, using an exact closure of the hierarchy obtained 
previously by Piasecki~\cite{pia_uno}. Such processes can model several physical 
situations as a recombination reaction in the gas phase or the fluorescence of 
laser excited gas atoms with quenching on contact (the one-dimensional aspect 
can be obtained by working in a suitable porous media~\cite{kopel}) or, the
annihilation of kink-antikink pairs in solid state physics~\cite{buti}.

We have recently investigated the case of the front formation for ballistic 
annihilation with two species initially separated in space in 
one-dimension~\cite{poisson}. This process can model, for example, the situation 
in which chemical species incorporated in a gel move ballistically under the 
action of a drift~\cite{zrini}. As the two species cannot penetrate one into the 
other (as they annihilate on contact), a well defined reaction front is formed. 
For a symmetrical Poissonian-type initial spatial distribution, we have 
proved~\cite{poisson} that the front behaves like a random walker for long time. 
In other words, at long time, the front can be anywhere on the line: it is 
delocalized.

In view of what has been done for the diffusive case, it is natural to try to 
find a system with ballistic-annihilation possessing a localized reaction front. 
Let us first remark that in the ballistic case, to impose a flux of particles at 
the boundaries of a finite system is equivalent to fix an initial spatial 
distribution of a large system. For example, the case with homogeneous spatial 
Poisson distribution can be mapped onto a system with boundary fluxes of
particles at constant rate, the time between two particles input 
being exponentially distributed.

Thus we shall consider the problem in terms of initial distribution of particles 
and ask the question: is it possible to localize the front by choosing a 
suitable initial distribution of the particles ? The answer is obviously yes. 
Indeed, as a simple example, consider the case in which initially each particle 
is located on a regular lattice, the $A$ particles being to the left of the 
origin and the $B$'s to the right, in a symmetric way. The position at which 
collision between one $A$ and one $B$ particle takes place defines the position 
of the annihilation front. Thus the front will always sit at the origin.
However, this case is pathological in the sense that the initial state is totaly 
ordered. For more general situations, with spatial randomness in the initial 
state, we have first to define what is meant by ''localisation of the front''. 
Then we shall show that indeed, it is possible to find suitable 
initial conditions localizing the front.

The paper is organized as follows. In section 2, we define precisely the class 
of models studied and introduce several criterions of localisation. In section 
3, we consider a class of correlated initial conditions for which we are able to 
compute the probability density $\mu(X;t)$ to find the front at position $X$ at 
time $t$ and which leads to the localization of the front. In section four, we 
consider a different class of initial conditions with no correlations between 
the initial positions of particles. For this more physical class of initial 
conditions, we are not able to compute explicitly $\mu(X;t)$, but we can show 
that the mean square position of the front $\langle  X^2 \rangle$ converges for 
$t\to\infty$. This means that the probability density $\mu(X;t)$ does not spread 
out. Remarks and conclusions are drawn in section 5.

\section{The model and the criterion of localization.}

We consider a one-dimensional system formed of two species of 
particles. Initially, particles $A$ are spatially randomly distributed in the 
region 
$(-\infty, 0)$ and the $B$ ones are spatially randomly distributed in the region 
$(0, \infty)$. For the sake of simplicity we shall consider only symmetric 
distributions in $A$ and $B$. In particular, $\rho_A(-x)= \rho_B(x)=\rho(x), 
x>0$, where $\rho_A$ and $\rho_B$ are the number densities of particles $A$ and 
$B$ respectively. For nonsymmetric distributions, the front moves acquiring some 
mean velocity.

The velocity of each particle is an independent random variable taking the  
value $\pm c$ with equal probability. Particles of the same kind suffer elastic 
collisions. When two particles $A$ and $B$ meet, they annihilate. Thus, 
practically the $A$ particles with velocity $-c$ and the $B$ particles with 
velocity $+c$ move freely. Accordingly, the relevant part of the 
dynamics concerns only the $A$ particles with velocity $+c$ and the $B$ ones 
with velocity $-c$.

Let $(y_1,y_2,\ldots,y_k,\ldots)$ be the initial positions of the $A$ particles 
with velocity $+c$ and $(x_1,x_2,\ldots,x_k,\ldots)$ the initial positions of 
the $B$'s with velocity $-c$. The particles are labeled such that 
$y_k<\cdots<y_1<0<x_1<\cdots<x_k$. The relative velocity between the $A$ and 
$B$ particles being $2c$, the pair of particles initially at $(x_k, y_k)$ will 
collide at time $t_k=(x_k-y_k)/2c$. The position at which this collision will 
take place defines the position of the annihilation front. Thus, the position of 
the front at time $t$ is $(x_1-y_1)/2c$, if $2ct\leq x_1+y_1$ (i.e. if no 
particle has yet collided), $(x_2-y_2)/2c$, if $x_1+y_1\leq 2ct\leq x_2+y_2$ 
(i.e. only the first particles on the right and on the left have collided), and 
so on. The dynamics in itself is purely deterministic, the only stochastic 
aspect comes from the initial conditions.

The properties of the front are completely defined by the probability density 
$\mu(X;t)$ to find the front at the point $X$, at time $t$. In a previous 
paper~\cite{poisson}, we have proved that: 
\begin{eqnarray}
\lefteqn{\mu(X;t) = 
    \left\langle \delta (X - \half[x_1+y_1])
    \right\rangle \label{eq:defmu} }\\
&+& \left\langle \sum_{k=1}^\infty
         \left[ \delta (X-\half[x_{k+1}+y_{k+1}]) - \delta (X-\half[x_k+y_k])
         \right] \theta (2ct-[x_k-y_k])
    \right\rangle \, , \nonumber
\end{eqnarray}
where $\delta$ and $\theta$ are the usual Dirac and Heaviside functions and the 
brackets denote the average over the initial positions, according to the 
initial distribution. It follows that, the second moment of $\mu(X;t)$ is
\begin{eqnarray}
\langle X^2 \rangle
&=& \left\langle \left( \frac{x_1+y_1}{2} \right)^2
    \right\rangle \nonumber \\
&+& \sum_{k=1}^\infty
    \left\langle
         \left[ \left( \frac{x_{k+1}+y_{k+1}}{2} \right)^2 -
                \left( \frac{x_k+y_k}{2} \right)^2
         \right] \theta \left( 2ct-[x_k-y_k] \right)
    \right\rangle\, . \label{eq:secmom}
\end{eqnarray}

Several criterions can be introduced to define the localization of the front.  
The first is simply to ask that $\mu(X;t)$ approches for $t \to \infty$ a 
limiting distribution with finite moments. A second criterion consists in
asking that $\mu(X;t)$ has a compact support with respect to $X$, for all time.  
These two criterions are useful provided that one is able to compute $\mu(X;t)$. 
This calculation will be possible for our first class of initial conditions 
(section 3).

However to deal with the second class of initial conditions, we are led to 
propose a weaker definition of the localization. The front will be said to be 
localized if $\mu(X;t)$ has a finite second moment (or variance in case of a 
non-symmetric initial distribution) for all time $t\leq\infty$.

We know what is happening in two limit cases. For a Poisson distribution the 
front is not localized while for particles regularly distributed on a lattice
the front is strongly localized. It is thus natural to consider an initial 
distribution which extarpolates smoothly between these two limits. Such 
distribution is provided by the Erlang-k-process~\cite{medhi}. For $k=1$, one 
recovers the Poisson process while for $k \to \infty$ one finds the case of the 
regular distribution. However, it can be shown~\cite{three} that for all finite 
values of $k$ the front is delocalized. Only the limit case $k \to \infty$ is 
localized.

\section{Correlated initial distribution.}

\begin{figure}[htb]
\epsfysize=8cm
\centerline{\epsfbox{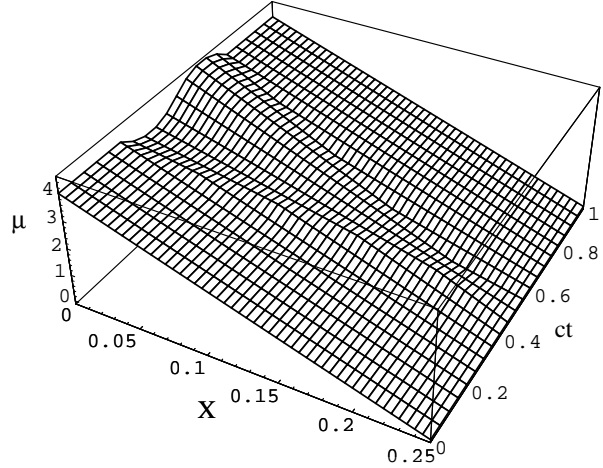}}
\caption{Plot of $\mu(X;t)$, for $\ell=\half$, $\delta=\quart$,
$X \in [0,\quart]$ and $ct \in [0,1]$. The probability density $\mu(X;t)$ is 
symmetric with respect to $X$ and periodic (with period $1$) with respect to 
$ct$.
\label{fig:one}}
\end{figure}

\begin{figure}[hbt]
\epsfysize=8cm
\centerline{\epsfbox{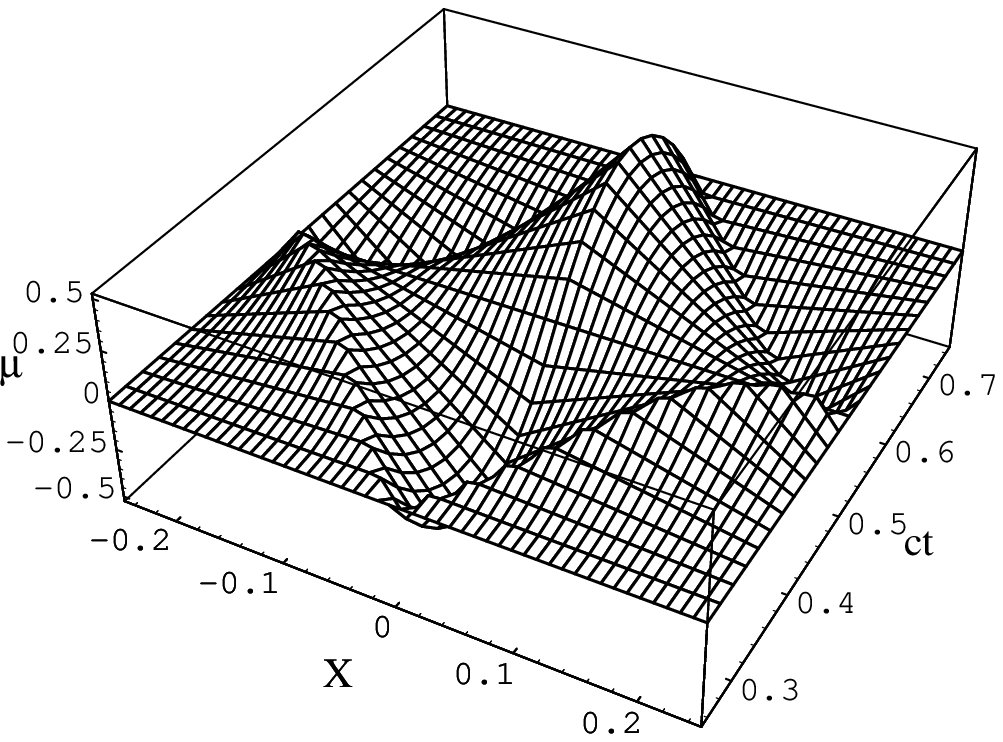}}
\caption{Plot of $\nu(X;t)$, for $\ell=\half$, $\delta=\quart$, $X 
\in [-\quart,\quart]$ and $ct \in [\quart, \frac{3}{4}]$.
\label{fig:two}}
\end{figure}

We aim at finding a class of initial conditions for which the probability 
density $\mu(X;t)$ has a compact support, with respect to X. This property can 
be obtained with the following initial condition: each $B$ particle is uniformly 
distributed in an interval of length $\delta < \ell$, centered 
at $(k-\half)\ell$, where $\ell$ is the distance (center-to-center) between two 
consecutive intervals and $k$ labels the particles $(k=1,2,\dots)$. We use the 
same distribution for particles $A$, but on the left of the origin. First, 
remark that the position of any particle is completely independent of the other; 
in addition we choose identical distribution for each particles, but at 
different regularly spaced positions. We eventually conclude that $\mu(X;t)$ 
must be a time periodic function of period $\ell/c$. Hence it is enough to 
compute $\mu(X;t)$ for $t\in[0,\ell/c)$. In this case, eq.~(\ref{eq:defmu}) 
simply reduces to
\begin{eqnarray}
\mu(X;t)
&\equiv&
    \nu_0(X) + \nu(X;t) \nonumber \\
&=& \left\langle \delta (X - \half[x_1+y_1]) \right\rangle \\
&+& \left\langle
         \left[ \delta (X - \half[x_2+y_2]) - \delta (X - \half[x_1+y_1])
         \right] \theta (2ct-[x_1-y_1])
    \right\rangle \, . \nonumber
\end{eqnarray}
The first bracket reads
\begin{eqnarray}
\nu_0(X)
&=& \frac{1}{\delta^2}
    \int_{-\frac{\ell+\delta}{2}}^{-\frac{\ell-\delta}{2}} {\d}y_1
    \int_{\frac{\ell-\delta}{2}}^{\frac{\ell+\delta}{2}} {\d}x_1 \,
    \delta (X-\half[x_1+y_1]) \nonumber \\
&=& \frac{4}{\delta^2} \left( \frac{\delta}{2}-|X| \right)
    \theta (\half\delta-|X|) \, .
\end{eqnarray}
The computation of the second bracket is  less straightforward, mainly due to 
the presence of the Heaviside function:
\begin{eqnarray}
\nu(X;t)
&=& \frac{1}{\delta^4}
    \int_{-\frac{3\ell+\delta}{2}}^{-\frac{3\ell-\delta}{2}} {\d}y_2
    \int_{\frac{3\ell-\delta}{2}}^{\frac{3\ell+\delta}{2}} {\d}x_2
    \int_{-\frac{\ell+\delta}{2}}^{-\frac{\ell-\delta}{2}} {\d}y_1
    \int_{\frac{\ell-\delta}{2}}^{\frac{\ell+\delta}{2}} {\d}x_1 \,
    \theta \left( 2ct-[x_1-y_1] \right) \nonumber \\
&\times&
    \left[ \delta (X-\half[x_2+y_2]) - \delta (X-\half[x_1+y_1]) \right]\, ,
\end{eqnarray}
However, simplifications arise by doing a Laplace-transform. Carrying the four
integrations over $y_1$, $y_2$, $x_1$ and $x_2$ and inverting, we finally 
obtain, $\nu(X;t)$ as a product of the Heaviside 
function $\theta(\half\delta-|X|)$ and a sum of five terms. We shall not 
reproduce these terms here as they are a little bit cumbersome. Nonetheless, we 
plotted both $\mu(X;t)$ and $\nu(X;t)$ for $t\in[0,\ell/c)$ (see 
figure~\ref{fig:one} and~\ref{fig:two}). The plot of $\nu(X;t)$ shows that it is 
not a simple function: already at this stage, $\mu(X;t)$ has a non-trivial 
shape.

It is clear that if we take an arbitrary distribution instead of a uniform one 
for the probability density inside each interval, we should always find that the 
support of $\mu(X;t)$ is $[-\half\delta,\half\delta]$, although the detailed 
shape of $\mu(X;t)$ may vary considerably. Moreover, one can easily generalize 
all these statements to non-symmetric distributions, by taking into account that 
the front will move. The support of $\mu$ will not be the same for all time, 
as it will follow the mean position of the front.

\section{Noncorrelated initial conditions.}

In this section we show that the front can be localized by a generalized  
Poisson-type distribution. For a homogeneous symmetric Poissonian particle 
distribution, we found~\cite{poisson}:
\begin{equation}
\langle X^2 \rangle \sim \frac{ct}{2\rho} + \mathrm{Const} \quad (t \to 
\infty)\, ,
\end{equation}
where $\rho$ is the initial density. Now suppose that instead of having a 
constant $\rho$, the density increases with position according to a power law:
\begin{equation}
\rho \equiv \rho(x) = x^\alpha \, . \label{alpha}
\end{equation}
Then, the front (whose average position is the origin) will see an increasing 
density in time. We can expect to obtain
\begin{equation}
\langle X^2 \rangle \sim \frac{(ct)^{1-\alpha}}{2} + \mathrm{Const} \quad
(t \to \infty)\, .
\end{equation}
For $\alpha>1$, $\langle X^2 \rangle$ converges to a constant, leading to a 
localization of the front, whereas for $\alpha<1$, it diverges and the front is 
delocalized. The mechanism leading to the this result is clear: when we increase 
the density, the mean distance between two successive particles decreases and 
thus two consecutive collisions should happen relatively often. As a 
consequence, as the time grows, the probability to find the front far from its 
mean value becomes smaller. We shall now make more precise this qualitative 
picture.

To begin we choose, as indicated, the initial particle distribution to be an 
inhomogeneous Poisson law. In other words, if there is a $B$ particle at 
position $x_1$, the probability to find its right nearest neighbor at position 
$x_2>x_1$ is
\begin{equation}
\pi(x_2,x_1) = \rho(x_2) \exp\left[-\int_{x_1}^{x_2}\rho(z){\d}z\right] \, ,
\label{eq:ipdf}
\end{equation}
where $\rho(x)$ should be understood as the initial particle density at point 
$x$. For $\rho(x)=\mathrm{Const}$, we find again the homogeneous Poisson law 
already studied. The probability to find $k-1$ particles between $0$ and $x_k$ 
and a particle at $x_k$ is
\begin{eqnarray}
\pi_k(x_k)
&=& \rho(x_k) \int_0^{x_k} {\d}x_{k-1}\,
    \exp\left[ -\int_{x_{k-1}}^{x_k} {\d}z\,\rho(z)\right] \nonumber \\
&\times&
    \rho(x_{k-1})\int_0^{x_{k-1}} {\d}x_{k-2}\,
    \exp\left[ -\int_{x_{k-2}}^{x_{k-1}} {\d}z\,\rho(z)\right] \cdots
    \nonumber \\
&\times&
    \rho(x_2) \int_0^{x_2} {\d}x_1\,
    \exp\left[ -\int_{x_1}^{x_2} {\d}z\, \rho(z)\right]
    \rho(x_1) \exp\left[ -\int_0^{x_1} {\d}z\,\rho(z)\right] \nonumber \\
&=& \rho(x_k) \exp\left[ -\int_0^{x_k} {\d}z\,\rho(z)\right]
    \frac{\left[ \int_0^{x_k} {\d}z\,\rho(z) \right]^{k-1}}{(k-1)!}\, .
\end{eqnarray}
$\pi_k$ is normalized to $1$ provided the density $\rho(x)$ does not vanish at 
infinity, which is satisfied for the law (\ref{alpha}), with $\alpha >0$. The 
distribution of particles A is chosen in a symmetric 
way $\rho_A(y)=\rho(-y),\,y<0$.

Using eq.~(\ref{eq:secmom}), the second moment becomes
\begin{eqnarray}
\langle X^2 \rangle_t
&=& \int_0^\infty {\d}x_1\,\rho(x_1) \exp\left( -\int_0^{x_1} \rho\right)
    \int_{-\infty}^0 {\d}y_1\,\rho(y_1) \exp\left( -\int_{y_1}^0 \rho\right)
    \left( \frac{x_1+y_1}{2} \right)^2 \nonumber  \\
&+& \sum_{k=1}^\infty 
    \int_0^\infty {\d}x_k\,\rho(x_k) \int_{-\infty}^0 {\d}y_k\,\rho(y_k)
    \int_{x_k}^\infty {\d}x_{k+1}\,\rho(x_{k+1})
    \int_{-\infty}^{y_k} {\d}y_{k+1}\,\rho(y_{k+1}) \nonumber \\
&\times&
    \exp\left( -\int_0^{x_{k+1}} \rho - \int_{y_{k+1}}^0 \rho\right)
    \frac{\left[ \int_0^{x_k} {\d}z\,\rho(z) \right]^{k-1}
          \left[ \int_{y_k}^0 {\d}z\,\rho(z) \right]^{k-1}}
         {\left[ (k-1)! \right]^2} \nonumber \\
&\times&
    \left[ \left( \frac{x_{k+1}+y_{k+1}}{2} \right)^2 -
           \left( \frac{x_k+y_k}{2} \right)^2
    \right] \theta \left( 2ct-[x_k-y_k] \right)\, ,\label{eq:mom}
\end{eqnarray}
where we used the notation
\[
\int_0^{x_k} \rho \equiv \int_0^{x_k} {\d}z\, \rho(z)\, .
\]
Now integrating twice by parts the expression
\begin{eqnarray*}
&&  \int_{x_k}^\infty {\d}x_{k+1}\, \frac{\partial}{\partial 
x_{k+1}}
    \left[ \exp\left( -\int_0^{x_{k+1}} \rho\right)\right] \\
&\times&
    \int_{-\infty}^{y_k} {\d}y_{k+1}\, \frac{\partial}{\partial 
y_{k+1}}
    \left[ \exp\left( -\int_{y_{k+1}}^0 \rho\right)\right]
    \left( \frac{x_{k+1}+y_{k+1}}{2} \right)^2
\end{eqnarray*}
and inserting the result into eq.~(\ref{eq:mom}), we obtain
\begin{eqnarray}
\langle X^2 \rangle_t
&=& \int_0^\infty {\d}x\,\rho(x) \int_0^\infty {\d}y\,\rho(y)
    \exp\left( -\int_0^x \rho - \int_0^y \rho\right)
    \left( \frac{x+y}{2} \right)^2 \nonumber  \\
&+& \sum_{k=0}^\infty \int_0^\infty {\d}x\,\rho(x) \int_0^\infty {\d}y\,\rho(y)
    \frac{\left( \int_0^x \rho \int_0^y \rho \right)^k}
         {\left( k! \right)^2} \nonumber \\
&\times&
    \left[ \rule{0mm}{7mm}
         \int_x^\infty {\d}x'
         \exp\left( -\int_0^{x'} \rho - \int_0^y \rho\right)(x'-y)
    \right. \nonumber  \\
&-& \left. \rule{0mm}{7mm}
         \half \int_x^\infty {\d}x' \int_y^\infty {\d}y'
         \exp\left( -\int_0^{x'} \rho - \int_0^{y'} \rho\right)
    \right] \theta \left( 2ct-[x+y] \right)\, .\label{eq:xsquare}
\end{eqnarray}
We go on by choosing:
\begin{equation}
\rho(x)=x^\alpha\, ,
\end{equation}
with $\alpha \ge 0$. As long as the time is finite, $\langle X^2 \rangle_t$ is 
finite for any $\alpha$, because the $\alpha=0$ case appears to be an upper 
bound. However it can grow arbitrarily when $t$ goes to infinity. Can we find 
$\alpha > 0$ such that $\langle X^2 \rangle_t$ is bounded for any time? It is 
very difficult to calculate exactly $\langle X^2 \rangle_t$. Even when $t$ is 
large, we have not been able to find its asymptotic expression. Nevertheless we 
can put $t\to\infty$ in the expression of $\langle X^2 \rangle_t$ and check its 
convergence. Clearly, as the first term of eq.~(\ref{eq:xsquare}) is time 
independent, we only have to care of the infinite sum. In addition, when 
$t\to\infty$, assuming that we can interchange the limit with the sum and the 
integrals, the restriction on the domain of integration introduced by the 
Heaviside function is removed. We are thus left to calculate
\begin{eqnarray}
a_k &\equiv&
    \int_0^\infty {\d}x \int_0^\infty {\d}y\, \frac{ x^\alpha y^\alpha}{(k!)^2}
    \left( 
\frac{x^{\alpha+1}}{\alpha+1}\frac{y^{\alpha+1}}{\alpha+1}\right)^k
    \nonumber \\
&\times&
    \left[
         \int_x^\infty {\d}x'\, (x'-y)
         \exp\left( -\frac{{x'}^{\alpha+1}}{\alpha+1}-
                  \frac{y^{\alpha+1}}{\alpha+1}
             \right)
    \right. \nonumber \\
&-& \left.\frac{1}{2} \int_x^\infty {\d}x' \int_x^\infty {\d}y' \,
    \exp\left(-\frac{{x'}^{\alpha+1}}{\alpha+1}
                  -\frac{{y'}^{\alpha+1}}{\alpha+1}
             \right)
    \right]
\end{eqnarray}
and to check the convergence of $\sum_{k=0}^\infty a_k$.

By changing the variables
\[
a=\frac{x^{\alpha+1}}{\alpha+1}\, ,\quad
b=\frac{y^{\alpha+1}}{\alpha+1}\, ,\quad
a'=\frac{{x'}^{\alpha+1}}{\alpha+1}\, ,\quad
b'=\frac{{y'}^{\alpha+1}}{\alpha+1}
\]
we get
\begin{eqnarray}
a_k
&=& \int_0^\infty {\d}a \int_0^\infty {\d}b\, \frac{(ab)^k}{(k!)^2}
    \left(\rule{0mm}{8mm}
         \int_a^\infty {\d}a'\, [(\alpha+1)a']^{-\frac{\alpha}{\alpha+1}}
         \e^{-a'-b}
    \right. \nonumber \\
&\times&
         \left\{ [(\alpha+1)a']^{-\frac{1}{\alpha+1}} -
                 [(\alpha+1)b]^{-\frac{1}{\alpha+1}}\right\} \nonumber \\
&-& \left.\rule{0mm}{8mm}
         \frac{1}{2} \int_a^\infty {\d}a' \int_b^\infty {\d}b'
         \left[ (\alpha+1)^2a'b' \right]^{-\frac{\alpha}{\alpha+1}}
         \e^{-a'-b'}
    \right) \, .
\end{eqnarray}
We finally obtain (by changing the order of integration over $a$ and $a'$ and 
over $b$ and  $b'$):
\begin{eqnarray}
a_k
&=& (\alpha+1)^{\frac{2}{\alpha+1}-1}
    \left\{\rule{0mm}{6mm}
         \frac{\Gamma\left( \frac{2}{\alpha+1}+k+1 \right)}{\Gamma(k+2)}
    \right. \nonumber \\
&-& \left.\rule{0mm}{6mm}
         \left[ k+1+\frac{1}{2(\alpha+1)} \right]
         \frac{\Gamma^2\left( \frac{1}{\alpha+1}+k+1 \right)}{\Gamma^2(k+2)}
    \right\}\, .
\end{eqnarray}
To study the convergence of $\sum_{k=0}^\infty a_k$, we simply have to study the 
asymptotic behavior of $a_k$. We find (see~\cite{abramo})
\[
a_k = \frac{1}{2} k^{-2+\frac{2}{\alpha+1}}
\left[ 1 - \frac{1}{\alpha+1} + \frac{2}{(\alpha+1)^2} +
       {\cal O}(k^{-1}) \right]\, ,
\]
when $k\to\infty$. It is thus evident that the sum converges if and only if 
$\int_{k=M}^\infty {\d}k\, k^{-2+\frac{2}{\alpha+1}}$ does (where $M$ is any 
integer sufficiently large). This leads us to the previously announced result: 
$\langle X^2 \rangle_t$ converges if $\alpha>1$, diverges if $\alpha<1$. For 
$\alpha=1$, one has a logarithmic divergence.

\section{Conclusion}

We have exhibited two different classes of initial conditions leading to a  
localization of the reaction front for this ballistic annihilation model. We 
have shown that the main physical reason for delocalization is a too slow decay 
of the  probability to find two consecutive particles at an arbitrarily large 
distance. Even an exponential law characterizing the Poisson distribution is not 
enough.

Several questions remain to be adressed. One natural extension of this work is 
to know whether or not a faster decay of the probability distribution of the 
nearest neighbours (for example a Gaussian probability instead of an exponential 
one) would lead to localization. Indeed, we know that if the interparticle 
distribution is
\[
\pi(x_2,x_1)=x_2\e^{-\half x_2^2 - \half x_1^2}
\]
(see eq.~(\ref{eq:ipdf})), then $\langle X^2 \rangle_t$ diverges 
logarithmically. 

Perhaps, the distribution
\[
\pi(x_2,x_1)=\e^{-\half (x_2-x_1)^2}\, ,
\] 
could lead to a front localization. Another possible extension is related to the 
case of a more general distribution of the velocities. These more difficult 
problems are under investigation.

\end{document}